\documentclass[floats,prb,aps,twocolumn]{revtex4}
\usepackage{amsmath}
\usepackage{graphicx}

\newcommand{\dprime}{{\prime\prime}}

\newcommand{\ef}{{\varepsilon_F}}
\newcommand{\slashi}{{i\kern-0.3em/}}
\newcommand{\slashone}{{1\kern-0.4em/}}
\newcommand{\sigmab}{{\overline \sigma}}
\begin{document}

\title{Analysis of the Disorder-Induced Zero Bias Anomaly in the Anderson-Hubbard Model}
\author{Hong-Yi Chen} 
\affiliation{Department of Physics, National Taiwan Normal University, Taipei 11677, Taiwan}
\author{R. Wortis}
\author{W. A. Atkinson} \email{billatkinson@trentu.ca}
\affiliation{Trent University, 1600 West Bank Dr., Peterborough ON,
  K9J 7B8, Canada} 
\date{\today}
\begin{abstract}
Using a combination of numerical and analytical calculations, we study
the disorder-induced zero bias anomaly (ZBA) in the density of states
of strongly-correlated systems modeled by the two dimensional
Anderson-Hubbard model.  We find that the ZBA comes from the response
of the nonlocal inelastic self-energy to the disorder potential, a result which
 has implications for theoretical approaches that retain only
the local self-energy.  Using an approximate analytic form for the
self-energy, we derive an expression for the density of
states of the two-site Anderson-Hubbard model.  Our formalism 
 reproduces the essential features of the ZBA, namely that the width
is proportional to the hopping amplitude $t$ and is independent of
the interaction strength and disorder potential.  
\end{abstract}
\maketitle

\section{Introduction.}
The Anderson-Hubbard model (AHM) is the simplest model that describes
strongly-correlated electrons in a disordered lattice.  The AHM is
widely used, for example, to describe doped transition metal oxides,
where the electronic properties are affected by both a strong local
Coulomb repulsion and doping-related disorder.\cite{Imada1998} The AHM
is also relevant to cold atomic gases in random optical
lattices,\cite{Schneider2008,White2009,Zhou2010} and there has been
recent interest in the AHM as a model interacting system that exhibits
Anderson
localization.\cite{Tusch1993,Heidarian2003,Byczuk2005,Kotlyar2001,Tanaskovic2003,Paris2007,Chakraborty2007,Song2008,Henseler2008,Henseler2009,Andrade2009,Byczuk2009,Pezzoli2010,Semmler2010}
The physics of the AHM is determined by dimensionality, by filling,
and by the three energy scales: the kinetic energy $t$, the on-site
Coulomb repulsion $U$, and the disorder strength $\Delta$.  When
$\Delta=0$, this model reduces to the well-known Hubbard model which,
despite its simplicity, has only been solved exactly in the limits of
one\cite{Lieb1968} and infinite dimensions\cite{Georges1996}.

In the Hubbard model, the interesting physics arises from a
competition between $t$, which tends to delocalize electrons, and $U$,
which tends to localize electrons.  When the lattice is half filled
(i.e.\ when there is one electron per site), a sufficiently large $U$
can generate a Mott insulating phase.  The Mott transition occurs at a
critical $U_c$ that depends on the details of the lattice.  Much of
the Hubbard model research in the past few decades has revolved around
strong correlation effects slightly away from the Mott insulating
phase, which is achieved either by taking $U$ less than $U_c$ or by
doping away from half filling.  One of the important ideas to come out
of the Hubbard model is that the low energy physics of the strongly
correlated metal phase near the Mott transition is governed by an
effective interaction $J\sim t^2/U$.\cite{Dagotto1994RMP}

\begin{figure}
\begin{center}
\includegraphics[width=\columnwidth]{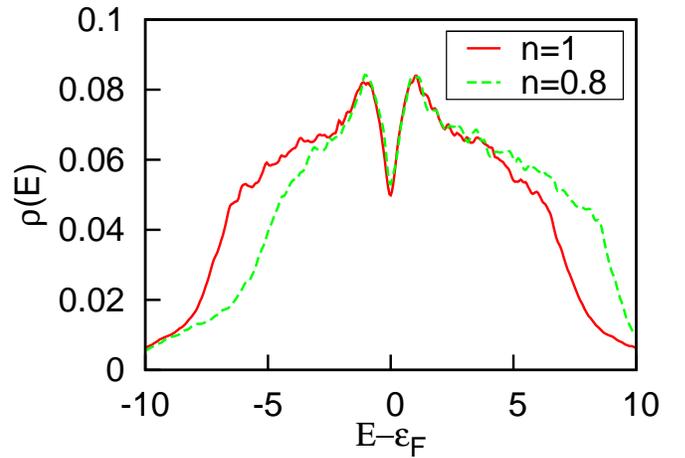}
\caption{(color online) Density of states for electron densities $n=1$
  (half filling) and $n=0.8$, showing the V-shaped zero bias anomaly
  at $\ef$.  Results are for exact diagonalization of 12 site
  lattices, and are averaged over 1000 disorder configurations.  Model
  parameters are $\Delta=20t$ and $U=8t$ throughout this work, unless
  stated otherwise.  }
\label{fig:dos}
\end{center}
\end{figure}

Contrary to this, recent exact diagonalization and quantum Monte Carlo
studies of the two-dimensional Anderson-Hubbard model have found that
a zero bias anomaly (ZBA) of width $t$ forms in the density of states
(DOS).\cite{Chiesa2008} The ZBA appears as
a V-shaped dip in the DOS at the Fermi energy $\varepsilon_F$, as
shown in Fig.~\ref{fig:dos}.  While it is not surprising that disorder
might introduce another low energy scale other than $t^2/U$, it is
surprising that this new scale is independent of both $U$ and
$\Delta$.

It is worth emphasizing that the observed ZBA is not explained by the
conventional Altshuler-Aronov theory of weakly correlated metals.  In
Altshuler-Aronov theory, the magnitude of the ZBA depends {\em
  inversely} on a dimension-dependent power of the Fermi
velocity,\cite{Altshuler1985} while the AHM ZBA grows linearly with the
Fermi velocity (which is approximately $2t$).

The physics of this ZBA is subtle, and is not captured by most
approximations. The Hartree-Fock
approximation\cite{Tusch1993,Heidarian2003,Fazileh2006,Chen2009,Shinaoka2009,Shinaoka2009b,Shinaoka2010}
yields a V-shaped zero bias anomaly when magnetic moments are allowed
to form,\cite{Chen2009} and has a low-energy soft gap that is apparently
associated with a multi-valley energy landscape.\cite{Shinaoka2009b}
However, the width of the ZBA grows with $U$, suggesting that
the physics of the ZBA is different than that found by exact
diagonalization. Furthermore, the evidence for a soft gap in exact
diagonalization calculations is less well
established,\cite{Shinaoka2009b} and it is possible that quantum
fluctuations fill in the soft gap. 
Another common approximation, dynamical mean field theory
(DMFT),\cite{Ulmke1995,Miranda2005,Laad2001,Balzer2005,Lombardo2006,Song2008,Andrade2009,Semmler2010}
includes strong correlation physics, but has not found a ZBA at
all. It has been argued\cite{Song2009} that this is because of
nonlocal contributions to the self-energy neglected in these
calculations.  Recent analytical studies of the two site AHM {\em do}
find a ZBA with qualitative features that are consistent with exact
diagonalization.  These calculations  interpret the ZBA in terms
of level repulsion between many-body
eigenstates,\cite{Wortis2010,Hongyi2010b,Wortis2010b} and demonstrate how strong
correlations can generate a kinetic energy driven ZBA.  While these
studies are instructive, it is difficult to connect them to the more
usual language of many-body self-energies in interacting systems.

In this article, we show how the ZBA arises from the response of the
inelastic self-energy to the disorder potential, using an approach
that is loosely based on one used by Abrahams et
al.\cite{Abrahams1981} to study the ZBA in weakly-correlated metals.
We restrict ourselves to two dimensions, where the existence of the
ZBA is well established, and work in the limit of strong disorder.  
In Sec.~\ref{sec:IIA}, we show that the ZBA comes from nonlocal
contributions to the local density of states, establishing (i) that
the ZBA is not a remnant of the Mott gap and (ii) that approximations
such as Hartree-Fock and DMFT (which retain only the local
self-energy) are missing key nonlocal physics.  In Sec.~\ref{sec:IIB},
we discuss an approximate self-energy, based on equation-of-motion
calculations,\cite{Song2009} which highlights the role of nonlocal
spin and charge correlations. We show numerically that this
approximation works well for large disorder, and then derive in
Sec.~\ref{sec:IIC} an approximate expression for the density of states
(DOS) based on this self-energy.  We find that the energy $t$ appears
as the natural energy scale for the ZBA.  The results are summarized
in Sec.~\ref{sec:III}.

\section{Calculations}
Before we proceed with the calculations, we emphasize a significant
difference between weakly and strongly correlated systems that affects
our analysis.  In the atomic limit, obtained by setting $t=0$, the DOS
is a sum of the local spectrum at each atomic site. For noninteracting
systems, each local spectrum has a single resonance at the orbital
energy $\epsilon$ of that site.  However, for strongly correlated
systems, there are two resonances, at $\epsilon$ and $\epsilon+U$,
which we term the lower Hubbard orbital (LHO) and upper Hubbard
orbital (UHO) respectively.  These energies correspond to transitions
in which an electron is added to a site that is initially empty (LHO)
or singly occupied (UHO).  The LHO and UHO are precursors of the lower
and upper Hubbard bands that form when $t$ is nonzero.

The calculations in this work are based on an expansion around the
atomic limit and are appropriate for the strong disorder case.  By
strong disorder, we mean $\Delta/2z \gg t$, where $z$ is the
coordination number of the lattice and $\Delta/2z$ is 
of the order of 
the average
level spacing of the $z$ sites adjacent to any site in the lattice,
and the factor of 2 is because there is an LHO and a UHO at each site.
In this limit, the local spectrum at a particular lattice site is
dominated by $2(z+1)$ resonances associated with the site and its $z$
nearest neighbors.\cite{Song2008}

\subsection{Analysis of Numerical Results}
\label{sec:IIA}
In this section, we develop a framework that explicitly shows the role
of local and nonlocal correlations in the DOS.  We then use this
framework to analyze the results of numerical exact diagonalization
calculations for the AHM.  We begin with a brief description of the exact
diagonalization calculations.

The AHM Hamiltonian is
\begin{equation}
\hat H = \sum_{i,j,\sigma} t_{ij} \hat c^\dagger_{i\sigma} \hat c_{j\sigma}
+ \sum_i \left (
\epsilon_i \hat n_i + U \hat n_{i\uparrow}\hat n_{i\downarrow}
\right ),
\label{eq:Ham}
\end{equation}
where $t_{ij}=-t$ for nearest-neighbor sites $i$ and $j$, and is zero
otherwise; $\hat c_{i\sigma}$ and $\hat n_{i\sigma}$ are the
annihilation and number operators for lattice site $i$ and spin
$\sigma$, and $\epsilon_i$ is the energy of the orbital at site $i$.
Disorder is introduced by choosing $\epsilon_i$ from a uniform
distribution $\epsilon_i \in
[-\frac{\Delta}{2},\frac{\Delta}{2}]$.

The AHM can be solved exactly for small clusters.  For our numerical
work, we use a standard Lanczos method\cite{Dagotto1994RMP} to find
the ground states of two-dimensional $N$-site ($N=10$, 12) clusters
with periodic boundary conditions, and then use a block-recursion
method to find the full nonlocal Green's function $G_{ij}(\omega)$ for
the lattice.\cite{Golub1996} The DOS is
\begin{equation}
\rho(E) = -\frac{1}{\pi N}\mbox{Im }\langle \sum_i  G_{ii}(E)\rangle,
\label{eq:rho0}
\end{equation}
  where $\langle \ldots \rangle$ indicates an average over disorder
  configurations at fixed chemical potential.  Examples of the
  disorder-averaged DOS are shown in Fig.~\ref{fig:dos}.

The goal of this section is to relate the DOS to two physically
interesting quantities, the local inelastic self-energy
$\Sigma_{ii}(\omega)$ and the nonlocal hybridization function
$\Lambda_{i}(\omega)$.  For a given disorder configuration, the
inelastic self energy is
\begin{equation}
{\bf \Sigma}(\omega) = {\bf G}_0(\omega)^{-1} -{\bf G}(\omega)^{-1},
\label{eq:exactSE}
\end{equation} 
where $(\ldots)^{-1}$ is a matrix inverse, ${\bf G}_0(\omega)$ is the
{\em noninteracting} Green's function for the same disorder
configuration as ${\bf G}(\omega)$, and $\Sigma_{ii}(\omega)$ is a
diagonal matrix element of ${\bf \Sigma}(\omega)$ (bold symbols
indicate matrices in the space of lattice sites).  The hybridization
function is then defined by
\begin{equation}
G_{ii}(\omega) = [\omega-\epsilon_i-\Sigma_{ii}(\omega) - \Lambda_{i}(\omega)]^{-1},
\label{eq:gloc}
\end{equation}
where $G_{ii}(\omega)$ is the local Green's function at site $i$, and
$\Sigma_{ii}$ is a diagonal matrix element of ${\bf \Sigma}(\omega)$.
Both $\Sigma_{ii}(\omega)$ and $\Lambda_{i}(\omega)$ can be extracted
from our numerical calculations: Eq.~(\ref{eq:exactSE}) gives
$\Sigma_{ii}(\omega)$, and then Eq.~(\ref{eq:gloc}) can be inverted to
find $\Lambda_i(\omega)$.

In the following analysis, we derive a formal expression for $\rho(E)$
in terms of $\Sigma_{ii}(\omega)$ and $\Lambda_i(\omega)$.  Our
starting point is Eq.~(\ref{eq:rho0}), with $G_{ii}(\omega)$ given by
Eq.~(\ref{eq:gloc}).  It is clear from these two equations that
$\rho(E)$ depends directly on $\Sigma_{ii}(\omega)$ and
$\Lambda_i(\omega)$, and the main issue we face in our derivation is
how to perform the disorder average in Eq.~(\ref{eq:rho0}).  We do
this in two steps: first, we take a partial disorder average of
$\Sigma_{ii}(\omega)$ and $\Lambda_i(\omega)$ over $\epsilon_j$ for
$j\neq i$ and for fixed $\epsilon_i$; second, we average
$G_{\epsilon_i}(\omega)$ over $\epsilon_i$.  As a result of the first
averaging process,
\begin{equation}
S_{i} \rightarrow S_{\epsilon}= \langle S_{i} \delta(\epsilon-\epsilon_i) \rangle
\label{eq:Sav}
\end{equation}
where $S_{i}(\omega) = \Sigma_{ii}(\omega) + \Lambda_{i}(\omega)$.   This gives the
average self-energy of all sites with energy $\epsilon$.  Then
\begin{equation}
G_{\epsilon}(\omega) \approx [\omega-\epsilon-S_{\epsilon}(\omega) ]^{-1}.
\label{eq:gloc2}
\end{equation}
Equation~(\ref{eq:Sav}) is the main approximation made in our
derivation, and we check below that we do not lose the physics of the
ZBA as a result of it.  The next step is to average
$G_{\epsilon}(\omega)$ over the local site energy.

To perform this average, we expand $S_\epsilon(\omega)$ about an
energy $E$ near $\varepsilon_F$, by analogy to what is done in Fermi
liquid theory.  In making this expansion, we consider two categories
of site: (i) sites with $\epsilon\sim E$ (LHO near $E$) and (ii) sites
with $\epsilon+U\sim E$ (UHO near $E$). Sites with neither $\epsilon$
nor $\epsilon+U$ near $E$ do not contribute to the DOS at $E$ and are
not included in our calculations.  For cases (i) and (ii)
\begin{eqnarray}
S_\epsilon(\omega) & \approx & S_{\overline E}(E) +
(\epsilon-\overline E)\partial_\epsilon S_\epsilon(E)_{\epsilon=\overline E}
\nonumber \\ 
&&
+ (\omega-E)
\partial_\omega S_{\overline E}(\omega)_{\omega=E} 
\label{eq:expandSE}
\end{eqnarray}
where $\overline E = E$ for case (i) and $\overline E = E-U$ for case (ii).
Then the local Green's function for site energy $\epsilon$ is
\begin{equation}
G_{\epsilon}(\omega) \approx \frac{Z}{\omega-E - (\epsilon-{\overline E})/m^\ast
-Z[S_{\overline E}(E)-\overline U]}
\label{eq:gloc3}
\end{equation}
with $Z = [1-\partial_\omega S_{\overline
    E}(\omega)]_{\omega=E}^{-1}$, ${m^\ast}^{-1} =
Z{[1+\partial_\epsilon S_\epsilon(E)]_{\epsilon={\overline E}}}$, and
$\overline U = E-\overline E$.  The final term in the denominator,
$S_{\overline E}(E) - \overline U$, vanishes identically in the atomic
limit (Appendix \ref{sec:D}).  Near the atomic limit,
$S_\epsilon(\omega)$ is complex, with small real and imaginary parts
that shift and broaden the orbital energies.  We show in
Sec.~\ref{sec:IIB} that the imaginary part of $S_\epsilon(E)$, which
results from disorder averaging, is of order $zt^2/\Delta$.
 
 Because the imaginary part of $S_\epsilon(E)$ is small, the average
of $G_\epsilon(E)$ over $\epsilon$ is easily done (Appendix~\ref{sec:C}), and we obtain the DOS
\begin{eqnarray}
\rho(E) &=& -\frac{\mbox{Im }}{\pi\Delta} \int_{-\Delta/2}^{\Delta/2}
G_\epsilon(E) d\epsilon \nonumber \label{eq:dos1} \\
&=& \frac{1}{\Delta}\left [  \frac{1}{1+\partial S_\mathrm{LHO}(E)}
+  \frac{1}{1+\partial S_\mathrm{UHO}(E)} \right ],
\label{eq:approxDOS}
\end{eqnarray}
where we have adopted the convenient notation
\begin{eqnarray*}
\partial S_\mathrm{LHO}(E) &\equiv& \mbox{Re }\partial_{\epsilon} S_{\epsilon}(E)|_{\epsilon=E} \\
\partial S_\mathrm{UHO}(E) &\equiv& \mbox{Re }\partial_{\epsilon} S_{\epsilon}(E)|_{\epsilon=E-U}.
\end{eqnarray*}
The two terms in the sum in Eq.~(\ref{eq:approxDOS}) give the partial
DOS for the LHO and UHO. For each term, there are two distinct
contributions: the first, $\partial_\epsilon \Sigma_\epsilon(E)$,
includes local Mott physics; the second,
$\partial_\epsilon\Lambda_\epsilon(E)$, includes the effects of
nonlocal self-energies.  Equation (\ref{eq:approxDOS}) is exact in
the atomic limit (Appendix \ref{sec:D}), and is a good approximation
for large disorder, where the imaginary part of $S_\epsilon(E)$ is
small and independent of $E$ (see Appendix \ref{sec:C}).  
equation is derived assuming $U<U_c$, where $U_c \approx \Delta$ in
the large disorder limit, since the system is a gapped Mott insulator
for $U>U_c$.

\begin{figure}[tb]
\includegraphics[width=\columnwidth]{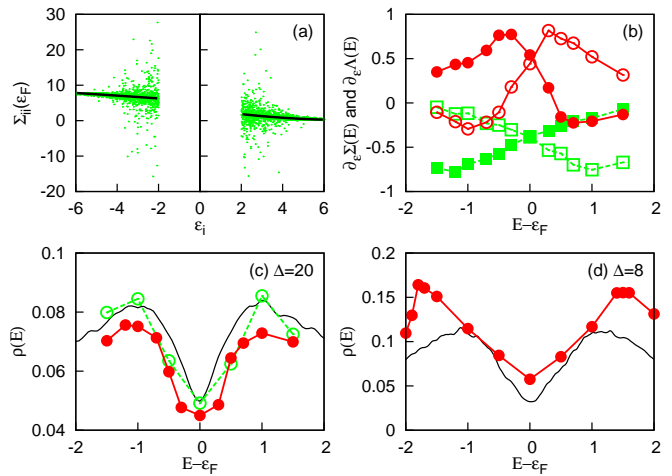}
\caption{(color online) Origin of the zero bias anomaly.  Data are for
  $n=1$, $U=8$, $\Delta=20t$, $t=1$, unless otherwise indicated.  (a)
  $\Sigma_{\epsilon_i}(E)$ for $E=\varepsilon_F=U/2$ versus site
  energy $\epsilon_i$.  Data is shown for sites with $\epsilon_i +U
  \approx E$ (left) and $\epsilon_i\approx E$ (right).  Lines are
  quadratic least-squares fits to the data in each region and are used
  to determine $\partial_\epsilon \Sigma_\epsilon(E)$; (b)
  $\partial_\epsilon \Sigma_\epsilon(E)|_{\epsilon=\overline E}$
  (squares) and $\partial_\epsilon
  \Lambda_\epsilon(E)|_{\epsilon=\overline E}$ (circles) for
  $\overline E=E$ (solid symbols) and $\overline E=E-U$ (empty
  symbols).  (c) The approximate DOS, calculated using the results
  from (b) in Eq.~(\protect\ref{eq:approxDOS}) (solid circles), the
  exact DOS (solid line), and the DOS for the approximate self-energy
  (\protect\ref{eq:approxSE}) (empty circles). (d) Results of a
  similar analysis for $\Delta=8$.}
\label{fig:ZBA}
\end{figure}

Equation~(\ref{eq:approxDOS}) gives an explicit relation between
$\rho(E)$ and the functions $\Sigma_\epsilon(\omega)$ and
$\Lambda_\epsilon(\omega)$.  One can interpret the derivatives
$\partial_\epsilon \Sigma_\epsilon(\omega)$ and $\partial_\epsilon
\Lambda_\epsilon(\omega)$ as the {\em response} of the self-energy and
hybridization function to changes in the local potential or,
equivalently, the response of these functions to the disorder
potential.  This is reminiscent of the situation in weakly correlated
metals, where a similar analysis related the ZBA to the response of
the charge density to the disorder potential.\cite{Abrahams1981}

We use Eq.~(\ref{eq:approxDOS}), in conjunction with our numerical
calculations, to establish the relative importance of
$\Lambda_i(\omega)$ and $\Sigma_{ii}(\omega)$ in forming the ZBA.  The
first step is to extract $\partial_\epsilon \Sigma_\epsilon(\omega)$
and $\partial_\epsilon \Lambda_\epsilon(\omega)$ from numerics.  This
process is illustrated in Fig.~\ref{fig:ZBA}.  For a given disorder
configuration, $\Sigma_{ii}(E)$ is calculated from
Eq.~(\ref{eq:exactSE}) at a fixed value of $E$, chosen to be $\ef$ in
Fig.~\ref{fig:ZBA}.  (Note that, for a finite size lattice, both
$\Lambda_i(\omega)$ and $\Sigma_{ii}(\omega)$ are real.) The collected
values of $\Sigma_{ii}(E)$ for all sites $i$ and for 1000
configurations are shown in Fig.~\ref{fig:ZBA}(a).
Data is shown for the two ranges, $\epsilon_i + U \approx
\varepsilon_F$ and $\epsilon_i \approx \varepsilon_F$, that contribute
to $\rho(\varepsilon_F)$.  A disorder-averaged $\Sigma_\epsilon(E)$ is
found by making least-squares quadratic fits to the data in each
range, from which the derivatives $\partial_\epsilon
\Sigma_\epsilon(\varepsilon_F)_{\epsilon=\varepsilon_F}$ and
$\partial_\epsilon
\Sigma_\epsilon(\varepsilon_F)_{\epsilon=\varepsilon_F-U}$ are
extracted. An identical set of calculations is then made for
$\partial_\epsilon \Lambda_\epsilon(\varepsilon_F)$.  The calculations
are repeated for other values of $E$, and resulting derivatives are
plotted as functions of $E$ in Fig.~\ref{fig:ZBA}(b).

As a check, we compare in Fig.~\ref{fig:ZBA}(c) the DOS from
Eq.~(\ref{eq:approxDOS}), calculated using the values shown in
Fig.~\ref{fig:ZBA}(b), with the exact DOS.  
The agreement between the two is very good.  We have repeated this
analysis for other values of $\Delta$, and continue to find
qualitative agreement down to the Mott transition at $\Delta \approx
U$ [Fig.~\ref{fig:ZBA}(d)].

Figure \ref{fig:ZBA}(b) shows the relative contributions to the ZBA
made by $\Sigma_{ii}(\omega)$ and $\Lambda_{i}(\omega)$.  The figure
shows that $\partial_\epsilon \Sigma_{\epsilon}(E)_{\epsilon=\overline
  E}$ is negative for both LHO ($\overline E=E$) and UHO ($\overline
E=E-U$).  From Eq.~(\ref{eq:approxDOS}), we see that a negative
derivative corresponds to an {\em increase} in $\rho(E)$, and not to
the V-shaped suppression of the DOS required to form a ZBA.
This result demonstrates that the ZBA does not come from the local
self-energy, and is therefore not a remnant of the Mott gap.  More
significantly, it demonstrates that the physics underlying the ZBA
cannot be reproduced by approximations that include only the local
self-energy, such as single-site DMFT or the Hartree-Fock
approximation.  The ZBA that appears in unrestricted Hartree-Fock
calculations must have a different origin than that found here.

In contrast to the self-energy derivative,
$\partial_\epsilon\Lambda_\epsilon(E)_{\epsilon=\overline E}$ is
positive for $E\approx \varepsilon_F$ and negative away from
$\varepsilon_F$, indicating that interorbital hybridization shifts
spectral weight {\em away} from $\varepsilon_F$.  This shows that the
ZBA comes from nonlocal correlations embedded in the hybridization
function.
On the one hand, this is not surprising since the Hubbard model in low
dimensions is known to map onto effective models with nonlocal
interactions; on the other hand, the energy scale $t$ of the ZBA is
not consistent with the energy scale $t^2/U$ of these effective
models.

We note one further interesting feature of Fig.~\ref{fig:ZBA}(b): the
plots of $\partial_\epsilon\Lambda_\epsilon(E)_{\epsilon= E}$ and
$\partial_\epsilon\Lambda_\epsilon(E)_{\epsilon=E-U}$ are asymmetric
with respect to $\varepsilon_F$.  This asymmetry indicates that LHOs
and UHOs behave differently when they are below or above
$\varepsilon_F$.  We will return to this point below.

In summary, we have established two main results in this section.
First, we have developed an expression, Eq.~(\ref{eq:approxDOS}), for
the DOS that relates $\rho(E)$ to the response of $\Sigma_\epsilon(E)$
and $\Lambda_\epsilon(E)$ to the disorder potential.  Second, we have
used this expression to analyze exact diagonalization results, and
have shown that the ZBA is the result of nonlocal correlations, rather
than the local self-energy.

\subsection{Structure of the Hybridization Function}
\label{sec:IIB}
In the previous section, we established that the ZBA can be related to
the derivative of $\Lambda_\epsilon(\omega)$ with respect to the site
energy $\epsilon$.  In this section, we analyse the structure of
$\Lambda_\epsilon(\omega)$ in more detail in order to see the role of 
spin and charge fluctuations in forming the ZBA.

We begin by writing $\Lambda_i(\omega)$ 
in terms of an alternative exact expression\cite{Georges1996} that is more
transparent than the original definition [Eq.~(\ref{eq:gloc})]:
\begin{equation}
\Lambda_{i}(\omega) = \sum_{j,k\neq i} [t_{ij}+\Sigma_{ij}(\omega)]
G_{jk}^{\slashi}(\omega)[t_{ki}+\Sigma_{ki}(\omega)],
\label{eq:lambda}
\end{equation}
where $G_{jk}^{\slashi}(\omega)$ is a Green's function matrix element
for the lattice with site $i$ removed.\cite{Gfn} This equation shows
explicitly how the matrix elements $t_{ij}+\Sigma_{ij}(\omega)$ couple
the site $i$ to the rest of the lattice.  In general,
$G_{jk}^{\slashi}(\omega)$ is not trivial to calculate, and this
expression is of use only when $G_{jk}^\slashi(\omega)$ can be
simplified through some approximation or limit.  Here, we are in the
limit of large disorder and low dimension, for which
$G_{jk}^{\slashi}(\omega)$ is approximately local.  In our discussion,
we thus consider only the dominant contributions, with $j=k$, in the
sum in Eq.~(\ref{eq:lambda}):
\begin{equation}
\Lambda_{i}(\omega) \approx \sum_{j\in \mathrm{nn}_i} [-t+\Sigma_{ij}(\omega)]^2
G_{jj}^{\slashi}(\omega),
\label{eq:lambda1}
\end{equation}
where  $j \in \mathrm{nn}_i$ indicates that $j$ is a  nearest neighbor of $i$. 

We note that, while $\Lambda_i(\omega)$ is real for a single disorder
configuration on a finite lattice, the disorder-averaged hybridization
function $\Lambda_{\epsilon}(\omega)$ is complex.  The real part of
$\Lambda_\epsilon(\omega)$ describes shifts of the LHO and UHO
energies while the imaginary part describes the broadening of these
orbitals due to the lattice.  For the analysis in this work to make
sense, the broadening must be much less than the level spacing
($\sim \Delta/2z$) of the local spectrum, so that discrete energy levels at
each site keep their distinct identity.  We can estimate the
broadening from a simplified disorder average of
Eq.~(\ref{eq:lambda1}).  Setting $\Sigma_{ij}=0$, we obtain
 \begin{eqnarray}
\Lambda^0_{\epsilon_i}(\omega) &=& zt^2 \langle G_{jj}^{\slashi}(\omega)\rangle_j,
\label{eq:lambda0}
\end{eqnarray}
where the sum over $j$ is replaced by the factor $z$, and $\langle
\ldots \rangle_j = \Delta^{-1} \int_{-\Delta/2}^{\Delta/2} \ldots
d\epsilon_j$ is the disorder average over site $j$.  This equation
assumes that $G_{jj}^\slashi(\omega)$ with different $j$ are
independent of each other.  The imaginary part of
Eq.~(\ref{eq:lambda0}) is
\begin{eqnarray}
\mbox{Im }\Lambda^0_{\epsilon_i}(\omega) &\approx& -\pi zt^2 \rho(\omega)  \sim -\frac{3z\pi t^2}{2\Delta},
\label{eq:lambda3}
\end{eqnarray}
 which gives a broadening of $O(zt^2/\Delta)$.  The condition that
 this is much less than the level spacing of the local spectrum can be
 written $2z^2t^2/\Delta^2 \ll 1$, which is met provided our initial
 assumption $2zt/\Delta \ll 1$ is met.

Equation (\ref{eq:lambda1}) shows that the nonlocal self-energy is
central to the ZBA.  To proceed further, we need an analytic form for
this self-energy, and we adopt a partial fractions expansion for the
self-energy that is based on the equation-of-motion
method.\cite{Song2009} The rationale for this choice is that the
equation-of-motion method correctly reproduces the LHO and UHO in the
atomic limit, and has been shown to be accurate for the two-site
AHM.\cite{Song2009} In general, we expect this method to work well
when short-range physics dominates.  The nonlocal self energy has the
form
\begin{eqnarray}
\Sigma_{ij}(\omega)=  
\frac{-tU^2p_{ij}}{(\omega-\epsilon_{i\sigma}-Uh_{i\sigmab})(\omega-\epsilon_{j\sigma}-Uh_{j\sigmab}) - \frac{\displaystyle O(t^2)}{\displaystyle \omega - \ldots}},\nonumber \\
\label{eq:approxSE}
\end{eqnarray}
where we suppress the explicit dependence of $\Sigma_{ij}(\omega)$ and $p_{ij}$ on $\sigma$
because we are considering only nonmagnetic phases, 
where
$h_{i\sigmab}=1-n_{i\sigmab}$, $n_{i\sigmab} = \langle \hat n_{i\sigmab}\rangle$,
with $\sigmab=-\sigma$, 
and where
\begin{equation}
p_{ij} = \langle \delta \hat n_{i\sigmab} \delta\hat n_{j\sigmab} \rangle + \langle \hat { S}_{i+} \hat {S}_{j-}\rangle -\langle \hat D_i^\dagger \hat D_j\rangle.
\end{equation}
(Here, $\langle \ldots \rangle$ indicates the expectation value,
rather than the disorder average.)  The three nonlocal correlations
making up $p_{ij}$ involve density fluctuation operators $\delta \hat
n_{i\sigma} = \hat n_{i\sigma} - n_{i\sigma}$, spin-flip operators
$\hat S_{i\pm}$, and pair annihilation operators $\hat D_{i} =
c_{i\downarrow}c_{i\uparrow}$.  The last of these three is an order of
magnitude smaller than the other terms and is discarded for the
remaining discussion.

In general, the usefulness of Eq.~(\ref{eq:approxSE}) is limited by
the difficulty of finding the higher-order terms in the continued
fraction.  These terms are important for determining the pole
structure of the self-energy, but do not change the fact that
$\Sigma_{ij} \propto p_{ij}$.  In the disorder-free Hubbard model, it
has been shown that these higher order terms 
are
qualitatively important;\cite{Odashima2005} however, the strongly
disordered case is close to the atomic limit and may be understood
qualitatively through a truncated self-energy, obtained by dropping
the $O(t^2)$ term in (\ref{eq:approxSE}).  We check this assertion
numerically: we calculate an approximate $\Lambda_{i}(\omega)$ using
the self-energy (\ref{eq:approxSE}) in Eq.~(\ref{eq:lambda1}), and
then calculate an approximate DOS using Eq.~(\ref{eq:approxDOS}).  The
results are plotted in Fig.~\ref{fig:ZBA}(c) in comparison with exact
diagonalization calculations, and the agreement between the two is
good.

We showed in the previous section that the ZBA comes from the response
of $\Lambda_\epsilon(E)$ to the disorder potential via the derivative
$\partial_{\epsilon} \Lambda_\epsilon(E)$.  The main idea suggested by
Eqs.~(\ref{eq:lambda1}) and (\ref{eq:approxSE}) is that this response is directly related to
the response of $\Sigma_{ij}(E)$, and therefore of $p_{ij}$, to the
disorder potential.  We show in the next section that there are other
contributions, but that a large part of the ZBA can indeed be traced
back to the response of the nonlocal charge and spin correlation
functions to the disorder potential.

We note that the form of $\Sigma_{ij}(\omega)$ explains the asymmetry
in $\partial_\epsilon\Lambda_\epsilon(E)_{\epsilon= E}$ and
$\partial_\epsilon\Lambda_\epsilon(E)_{\epsilon=E-U}$ with respect to
$\ef$, shown in Fig.~\ref{fig:ZBA}(b).  This figure shows that the ZBA
is formed from a shift away from $\ef$ of LHOs below $\ef$ and of UHOs
above $\ef$.  According to Eq.~(\ref{eq:approxSE}), this asymmetric
shift occurs because the correlation $p_{ij}$ is largest when sites
$i$ and $j$ are both singly-occupied, namely when $\ef-U \lesssim
\epsilon_i, \epsilon_j \lesssim \ef$. 
(The spin correlations vanish
when either site is empty or doubly occupied.)  
This condition on
$\epsilon_i$ and $\epsilon_j$ is equivalent to the requirement, at
each site, that the LHO be below $\ef$ and the UHO be above $\ef$.

In summary, we have used a form for the hybridization function that
shows explicitly the role of the nonlocal self-energy.  We have
proposed using an analytic form, Eq.~(\ref{eq:approxSE}), for this
self-energy, and have shown numerically that it reproduces the density
of states obtained by exact diagonalization.  The main result of this
section is that the nonlocal self-energy, and therefore the ZBA, depends
on nonlocal spin and charge correlations. 

Ideally, one would now like to use this
formalism to derive an analytic expression for the density of
states; this requires knowledge of $p_{ij}$ and is in general quite
difficult since $p_{ij}$ is different along every bond in the lattice.
In the next section, we therefore focus on a simple
model for which $p_{ij}$ is known, and the DOS can be found analytically.

\subsection{Density of States}
\label{sec:IIC}
As a simple application of the formalism derived in the previous
sections, we calculate the DOS for the two-site AHM (2SAHM).  This
model has been studied elsewhere by direct diagonalization of the
Hamiltonian,\cite{Wortis2010,Hongyi2010b,Wortis2010b} and provides a point of
comparison for the current work.  Our approach is straightforward: we
use the self-energy (\ref{eq:approxSE}) to find an approximate
hybridization function with which we evaluate the density of states
using Eq.~(\ref{eq:approxDOS}).

The 2SAHM consists of an ensemble of two-atom ``molecules'' with random
site energies $\epsilon_i$ and $\epsilon_j$.
 The disorder averaged
hybridization function for site $i$ is
\begin{equation}
\Lambda_{\epsilon_i}(\omega) = \langle   [-t+\Sigma_{ij}(\omega)]^2
G_{jj}^{\slashi}(\omega)\rangle_j.
\label{eq:lambda2}
\end{equation}
In this form, the hybridization function has a useful symmetry
(Appendix \ref{sec:B})
\begin{equation}
\partial\Lambda_\mathrm{UHO}(\ef+\tilde E) = \partial\Lambda_\mathrm{LHO}(\ef-\tilde E),
\label{eq:sym}
\end{equation}
where $\partial \Lambda_\mathrm{LHO}(E)$ and $\partial \Lambda_\mathrm{UHO}(E)$ have similar
definitions as $\partial S_\mathrm{LHO}(E)$ and $\partial S_\mathrm{UHO}(E)$, and
$\tilde E$ is the energy $E$ measured relative to $\ef$.

One consequence of this symmetry is that contributions to
$\partial\Lambda_\mathrm{LHO}(E)$ that are even under $\tilde E\rightarrow
-\tilde E$ are more important for the ZBA than those which are odd.
To show this, we define $\delta \rho(E)$ to be the change in the DOS
due to the hybridization function, namely $\delta \rho(E) = \rho(E) -
\rho_0(E)$, where $\rho_0(E)$ is evaluated with
$\partial_\epsilon\Lambda_\epsilon(E)$ set to zero.  To linear order
in $\partial_\epsilon \Lambda_\epsilon(E)$, Eq.~(\ref{eq:approxDOS})
gives
\begin{eqnarray}
\delta \rho(E) &=&  -\frac{1}{\Delta} \left [
 \frac{\partial  \Lambda_\mathrm{LHO} (E) }{[1+\partial \Sigma_\mathrm{LHO}(E)]^2}
 +
  \frac{\partial  \Lambda_\mathrm{UHO} (E) }{[1+\partial \Sigma_\mathrm{UHO}(E)]^2}
  \right]. \nonumber \\
\end{eqnarray}
Noting, from Fig.~\ref{fig:ZBA}(b), that
$\partial\Sigma_\mathrm{LHO}(E)\approx \partial
\Sigma_\mathrm{UHO}(E)$ near $\ef$, we get
\begin{equation}
\delta \rho(E) \sim -\frac{1}{\Delta} \frac{\partial \Lambda_\mathrm{LHO}(E)+\partial\Lambda_\mathrm{UHO}(E)}{[1+\partial \Sigma(\ef)]^2}.    
\label{eq:iffyDOS}
\end{equation}
From this, and from Eq.~(\ref{eq:sym}), it follows that the most
significant contributions to $\delta \rho(E)$ come from terms in
$\partial \Lambda_\mathrm{LHO}(E)$ that are even in $\tilde E$.

To calculate $\partial \Lambda_{\mathrm{LHO}}(E)$, we expand
Eq.~(\ref{eq:lambda2}) as $\Lambda_{\epsilon}(\omega) =
\Lambda^0_{\epsilon}(\omega)+\Lambda^\prime_\epsilon(\omega)+\Lambda^\dprime_\epsilon(\omega)$,
where 
\begin{eqnarray}
\Lambda^0_{\epsilon_i}(\omega) &=& t^2 \langle G^\slashi_{jj}(\omega) \rangle_j\\
\Lambda^\prime_{\epsilon_i}(\omega) &=&  -2t \langle \Sigma_{ij}(\omega) G^\slashi_{jj}(\omega)\rangle_j \\
\Lambda^\dprime_{\epsilon_i}(\omega) &=& \langle \Sigma_{ij}(\omega)^2 G^\slashi_{jj}(\omega)\rangle_j.
\end{eqnarray}
We have evaluated each of these terms analytically and find that, by far, the
largest contribution to the ZBA comes from $\partial
\Lambda^\dprime_{\mathrm{LHO}}(E)$.  In particular, $\partial
\Lambda^0_\mathrm{LHO}(E)$ is an odd function of $\tilde E$ and
therefore makes almost no contribution to the ZBA; $\partial
\Lambda^\prime_\mathrm{LHO}(E)$ contains both odd and even terms and
therefore does contribute to the ZBA, but is an order of magnitude
smaller than $\partial \Lambda^\dprime_\mathrm{LHO}(E)$.  It is,
perhaps, not surprising that the term containing the highest power of
$\Sigma_{ij}(\omega)$ makes the largest contribution to the ZBA.  For
clarity, we include only results for $\Lambda^\dprime_{\epsilon_i}(E)$
in our calculation of $\delta \rho(E)$.

\begin{widetext}
Using
Eq.~(\ref{a:aG2}) for $G_{jj}^\slashi(\omega)$, we obtain
\begin{eqnarray}
\Lambda^\dprime_{\epsilon_i}(\omega)  &=& \frac{t^2U^4}{\Delta}\int_{-\Delta/2}^{\Delta/2} d\epsilon_j 
\frac{p_{ij}^2}{(\omega-\epsilon_i-Uh_{i\sigmab})^2(\omega-\epsilon_j-Uh_{j\sigmab})(\omega-\epsilon_j)(\omega-\epsilon_j-U)},
\label{eq:Lexpand}
\end{eqnarray}
and differentiating this with respect to $\epsilon_i$, we obtain
\begin{eqnarray}
\frac{\partial \Lambda_{\epsilon_i}^\dprime(\omega)}{\partial \epsilon_i} &=& 
\frac{t^2U^4}{\Delta} \int_{-\Delta/2}^{\Delta/2} d\epsilon_j  \left [
\partial_{\epsilon_i}p_{ij}^2
+\frac{2p_{ij}^2(1+U\partial_{\epsilon_i}h_{i\sigmab})}{(\omega-\epsilon_i-Uh_{i\sigmab})}
+\frac{p_{ij}^2U\partial_{\epsilon_i}h_{j\sigmab}}{(\omega-\epsilon_j-Uh_{j\sigmab})}
\right ] \nonumber \\
&&\times\frac{1}{(\omega-\epsilon_i-Uh_{i\sigmab})^2(\omega-\epsilon_j-Uh_{j\sigmab})(\omega-\epsilon_j)(\omega-\epsilon_j-U)}
\label{eq:Lderivative}
\end{eqnarray}
To calculate $\partial \Lambda^\prime_\mathrm{LHO}(E)$, we set
$\omega=\epsilon_i=E$ in Eq.~(\ref{eq:Lderivative}).  Then there are
four terms, proportional to $\partial_{\epsilon_i} p_{ij}^2$, to
$p_{ij}^2\partial_{\epsilon_i} h_{i\sigmab}$, to
$p_{ij}^2\partial_{\epsilon_i} h_{j\sigmab}$, and to $p_{ij}^2$.  The
last of these is a factor $t/U$ smaller than the others and is
discarded.  
\end{widetext}

Because of the simplicity of the 2SAHM, we can write the coefficients
$p_{ij}$, $h_{i\sigmab}$, and $h_{j\sigmab}$ in terms of the many-body
wavefunction for the two site system, and thus find their explicit
dependence on $\epsilon_i$ and $\epsilon_j$.  This makes the
integration over $\epsilon_j$ possible.  The calculations are
complicated by the fact that we do the integration at fixed chemical
potential, meaning that the number of electrons in the ground state
depends on $\epsilon_i$ and $\epsilon_j$. The dominant contribution to
the ZBA comes from cases where the ground state has two electrons, and we
include only this term in our result. The calculations are 
somewhat lengthy, and we leave the details to Appendix \ref{sec:A}.

The result of these calculations is, from Eq.~(\ref{eq:iffyDOS}) and
Eq.~(\ref{a:dosdprime}),
\begin{eqnarray}
\delta \rho(E) & \approx &
  \frac{-28\sqrt{2}t}{27\Delta^2(1+\partial \Sigma)^2}
 \left [ F_2(x) + F_4(x) 
+ \frac {3\pi}{ 4} \right ]
 \label{eq:theorydos}
\end{eqnarray}
where 
$x = (2t^2-\tilde E^2)/(2\sqrt{2}t|\tilde E|)$,
with $\tilde E = E-\ef$ and 
\begin{eqnarray*}
F_2(x) &=&  \frac{x}{x^2+1}+\tan^{-1}(x), \\
F_4(x) &=&  \frac 12 F_2(x) - \frac{x}{(x^2+1)^2}.
\end{eqnarray*}

\begin{figure}
\includegraphics[width=\columnwidth]{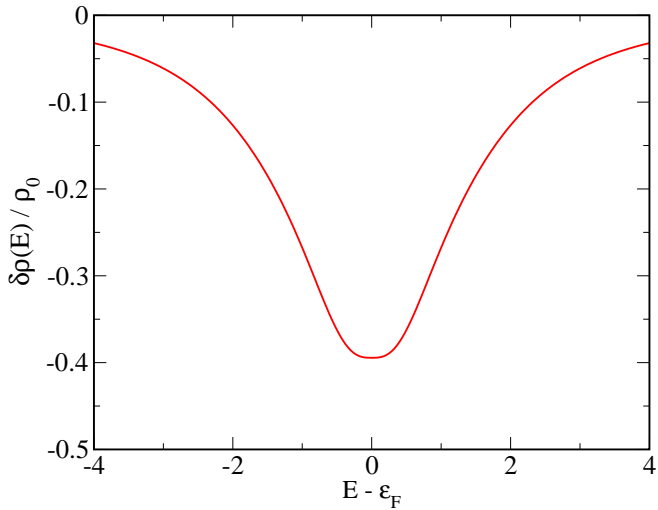}
\caption{Theoretical density of states from
  Eq.~(\protect\ref{eq:theorydos}).  Results are shown for $\Delta/t =
  20$. Note that the curve is independent of $U$.}
\label{fig:two-site}
\end{figure}

Equation (\ref{eq:theorydos}) is plotted in Fig.~\ref{fig:two-site}
for the case $\Delta=20t$.  For this plot, the unknown prefactor
$1+\partial \Sigma$ is taken to be $0.7$, based on the value of
$\partial_\epsilon \Sigma_\epsilon(E)|_{\epsilon=\ef}$ in
Fig.~\ref{fig:ZBA}(b).  The resulting plot is qualitatively consistent
with exact results for the 2SAHM\cite{Wortis2010,Hongyi2010b}; from Eq.~(\ref{eq:theorydos}), the width of the ZBA is of order $2\sqrt{2}t$, and the depth is
proportional to $t/\Delta^2$.

In previous studies of the 2SAHM, the ZBA was attributed to level
repulsion between many-body states.  Here, level repulsion is implicit
in $\partial \Lambda_\mathrm{LHO}(E)$ and $\partial
\Lambda_\mathrm{UHO}(E)$, since these describe the shifts of the
atomic LHO and UHO due to neighboring sites.  These
shifts are primarily due to the response $\partial_{\epsilon_i}
\Sigma_{ij}(E)$ of the nonlocal self-energy to the disorder potential.
In Eq.~(\ref{eq:Lderivative}), we showed that $\partial_{\epsilon_i}\Sigma_{ij}(E)$
depends on the local charge susceptibilities $\partial_{\epsilon_i}
h_{i\sigmab}$ and $\partial_{\epsilon_i} h_{j\sigmab}$, and a
generalized susceptibility $\partial_{\epsilon_i} p_{ij}$.  For the
2SAHM, the last term is the largest, so that the ZBA is mostly due to
the response of the nonlocal spin and charge correlation functions
making up $p_{ij}$.

It is interesting to note that Mott physics suppresses this response.
This is because the local Coulomb interaction tends to fix the charge
density at each site, so that the spin and charge correlations are
only weak functions of $\epsilon_i$.  For example, configurations in
which $\epsilon_i$ and $\epsilon_j$ are near $\ef$ have a singlet
ground state $|s\rangle$, with corrections of order $t/U$.  A small
change in $\epsilon_i$ changes this ground state, and therefore
$p_{ij}$, by order $t/U$.  Thus $\partial_{\epsilon_i}p_{ij}$ is
suppressed by Mott physics.  This is not the case when $\epsilon_i$
and $\epsilon_j+U$ are within $t$ of $\ef$.  Then $|s\rangle$ and
$|02\rangle$ are nearly degenerate, and the proportions of $|s\rangle$
and $|02\rangle$ making up the ground state vary linearly with
$\epsilon_i$.  In this regime, $\partial_{\epsilon_i} p_{ij}$ is not
small.  The ZBA therefore comes from disorder configurations in which
Mott physics does not suppress nonlocal charge fluctuations.

The results presented in this section are valid for $\Delta \gtrsim U
\gg t$.  When $U\gtrsim \Delta$, the spectrum has distinct lower and
upper Hubbard bands.  In our calculations for the 2SAHM, the ZBA
collapses rapidly when the Hubbard bands no longer overlap, since
configurations with degenerate LHO and UHO no longer occur.  This
appears to contradict results reported by Chiesa et
al.,\cite{Chiesa2008} where the ZBA persisted for $U>\Delta$, away
from half-filling.  Direct comparison with
Ref.~\onlinecite{Chiesa2008} is not straightforward since they are not
in the regime $\Delta \gg zt$ in which our theory is valid.  We have
performed preliminary exact diagonalization calculations for one- and
two-dimensional clusters for the case $U>\Delta \gg zt$; these show
that while the {\it slope} of the ZBA (namely, $\partial_E\rho(E)$) is
approximately independent of $U$, the {\it width} and {\it depth} are
stronger functions of $\Delta$ than when $U<\Delta$.  We find that the
width of the ZBA is not simply $t$ in the gapped phase; however, these
results are preliminary, and a careful study is required to resolve
this discrepancy.

\section{Conclusions}
\label{sec:III}

In this work, we have discussed the origins of the disorder-induced
zero bias anomaly in the Anderson-Hubbard Model.  Several aspects of
this zero bias anomaly are unique to strongly correlated systems with
short range interactions.  Most significant is the fact that the width
of the anomaly is set by the hopping matrix element $t$, and is
independent of the interaction strength $U$ and disorder potential
$\Delta$ over a wide range of $\Delta$ and $U$.  In the two-site
Anderson-Hubbard model, this has been understood as the result of
level repulsion between lower and upper Hubbard
orbitals.\cite{Wortis2010,Hongyi2010b}

Here, we have gone beyond the 2SAHM, and have shown that the underlying physics of
the zero bias anomaly in larger
clusters can be extracted from an analysis of exact diagonalization
calculations.  The analysis is based on an expansion around the atomic
limit, and is appropriate for disorder $\Delta$ much larger than the
clean-limit bandwidth $zt$.  Through this analysis, we have found that
the local Coulomb interaction generates nonlocal spin and charge
correlations between adjacent lattice sites, which cause an overall
shift of spectral weight away from the Fermi energy $\ef$.  By this
mechanism, a V-shaped zero bias anomaly is formed in the density of
states at $\ef$.

Specifically, the zero bias anomaly comes primarily from the response
$\partial_{\epsilon_i}\Sigma_{ij}(E)$ of the nonlocal self-energy to
the disorder potential.  Mott physics tends to suppress this response;
however, disorder configurations in which many-body Fock states are
nearly degenerate are sensitive to small changes in the lattice
potential, and for these configurations
$\partial_{\epsilon_i}\Sigma_{ij}(E)$ is not small.

Using the formalism developed in this work, we have obtained an
analytic expression for the DOS of a two-site Anderson-Hubbard model.
This expression reproduces the essential physics of the zero bias
anomaly found numerically; the anomaly has a width of order $t$, and a
depth which is independent of $U$ when $U\gg t$.

\section*{Acknowledgments}
We acknowledge support by NSERC, CFI and OIT.  This work was made possible by the facilities of the Shared Hierarchical Academic Research Computing Network (SHARCNET) and the High Performance Computing Virtual Laboratory (HPCVL). H.-Y.C.\ is supported by NSC Grant No.\ 98-2112-M-003-009-MY3.

\appendix
\section{Results for the Atomic Limit}
\label{sec:D}
In the atomic limit, the exact local Green's function in the nonmagnetic phase is
\[
G_i(\omega) = \frac{1}{\omega-\epsilon_i -\Sigma_{ii}(\omega)}
\]
where the exact self-energy is,\cite{Song2009} 
\begin{equation}
\Sigma_{ii}(\omega) = U\frac{n_i}{2} 
+ \frac{U^2\frac{n_i}{2}(1-\frac{n_i}{2})}{\omega-\epsilon_i-U(1-\frac{n_i}{2})},
\label{a:SEloc}
\end{equation}
and the charge density is 
\begin{equation}
n_{i} = \left \{ \begin{array}{lr} 
2, & \epsilon_i + U < \ef \\ 
1, & \ef-U < \epsilon_i < \ef \\
0 & \epsilon_i > \ef
\end{array}\right  ..
\label{a:n}
\end{equation}
First, we note that it follows directly from (\ref{a:SEloc}) and (\ref{a:n}) that the term $S_{\overline E}(E) - \overline U$ in Eq.~(\ref{eq:gloc3}) vanishes identically in the atomic limit.  

Next, we check that Eq.~(\ref{eq:approxDOS}) is exact in the atomic limit.  
From Eq.~(\ref{a:SEloc}), 
\begin{eqnarray}
\left .\partial_{\epsilon_i} \Sigma_{ii}(E) \right |_{\epsilon_i = E} & = & \frac{n_i/2}{1-n_i/2} \\
\left .\partial_{\epsilon_i} \Sigma_{ii}(E) \right |_{\epsilon_i = E-U} & = & \frac{1-n_i/2}{n_i/2} 
\end{eqnarray}
Taking, for example, $E$ slightly less than $\ef$ we obtain
\begin{eqnarray}
\rho(E) &= &\frac{1}{\Delta}\left [ 
\left. \frac{1}{1 + \partial_\epsilon \Sigma_\epsilon(E)}\right |_{\epsilon=E} +
\left. \frac{1}{1 + \partial_\epsilon \Sigma_\epsilon(E)}\right |_{\epsilon=E-U}\right ] \nonumber \\
&=& \frac{1}{\Delta} \left [ \frac{1}{1+1} + \frac{1}{1+0} \right ] \nonumber\\
&=& \frac{3}{2\Delta},
\end{eqnarray}
for the disorder-averaged DOS.  This is the exact result.

\section{Density of States for Complex Self Energies}
\label{sec:C}

If the self-energy $S_\epsilon(E)$ is complex, then the analysis leading to Eq.~(\ref{eq:gloc3}) is unchanged,  
\begin{equation}
G_{\epsilon}(E) \approx \frac{-1}{ (\epsilon-{\overline E})[1+\partial S]
+[S-\overline U]},
\end{equation}
where we use the compact notation $\partial S = \partial_\epsilon S_\epsilon(E)|_{\epsilon=\overline E}$ and $S = S_{\overline E}(E)$; however,
the disorder-averaged density of states is
\begin{eqnarray}
\rho(E) 
&\approx& \frac{1}{\pi\Delta} \mbox{Im}\sum_{\overline E = E,E-U}
\frac{1}{1+ \partial S} \times
\nonumber \\
&&  \ln \left[ \frac{ (\frac{\Delta}{2}-\overline E)(1+ \partial S) + S-\overline U } { (-\frac{\Delta}{2}-\overline E)(1+ \partial S) + S-\overline U } \right]  ,
\label{a:dos1}
\end{eqnarray}
where the argument of the logarithm is complex.  $E$ has an infinitessimal positive imaginary part so that  Eq.~(\ref{a:dos1}) reduces to Eq.~(\ref{eq:dos1}) when $S$ and $\partial S$ are real.  

In this work, $S_{\overline E}(E)$ is complex as a result of the disorder averaging process.  We find that the hybridization function introduces imaginary components,
\begin{eqnarray}
S&\rightarrow& S - i\Gamma \\ 
\partial S&\rightarrow &\partial S + i\gamma
\end{eqnarray}
where $\Gamma \sim zt^2/\Delta$ and $\gamma \sim zt/\Delta$.  Near the Fermi energy at half filling, $E\sim U/2$ such that 
\begin{eqnarray*}
\left |(\pm \frac {\Delta}{ 2} - E)(1+\partial S)\right | &\gg& |S-\overline U|, \\
\left |(\pm \frac {\Delta}{ 2} - E)\right | \gamma &\gg& \Gamma,
\end{eqnarray*}
except near the Mott transition at $U\approx \Delta$.
Then 
\begin{eqnarray}
\rho(E) 
&\approx& \frac{1}{\pi\Delta} \sum_{\overline E = E,E-U} \Bigg \{
\frac{1}{1+ \partial S} \left [ \tan^{-1}\frac{\gamma}{1+\partial S}
\right. \nonumber \\
&& \left . 
- \tan^{-1} \frac{-\gamma}{-(1+\partial S)}\right ] 
-\frac{\gamma}{1+\partial S} \ln \left| \frac{ (\frac{\Delta}{2}-E)} { (\frac{\Delta}{2}+E) } \right|  \Bigg \}, \nonumber \\
&\approx& \frac{1}{\pi\Delta} \sum_{\overline E = E,E-U} \Bigg \{
\frac{\pi}{1+ \partial S} 
-\frac{\gamma}{1+\partial S} \ln \left| \frac{ \frac{\Delta}{2}-E} { \frac{\Delta}{2}+E } \right|  \Bigg \},\nonumber \\
\label{a:dos2}
\end{eqnarray}
where $\tan^{-1} [-\gamma/-(1+\partial S)] \approx -\pi+\gamma/(1+\partial S)$.
The first term in Eq.~(\ref{a:dos2}) is the result found in Eq.~(\ref{eq:dos1}), while
the second term increases $\rho(E)$ by order $zt/\Delta^2$.  This term is comparable
in magnitude to the corrections responsible for the ZBA, but is featureless near $E=\ef$, and therefore does not contribute to the ZBA.  The conclusion to be drawn from this appendix is that the expression (\ref{eq:dos1}) is sufficient to understand the ZBA provided $zt/\Delta\ll 1$.

\section{Derivation of  $\partial \Lambda^\dprime_\mathrm{LHO}(\omega)$}
\label{sec:A}
 
\begin{figure}
\begin{center}
\includegraphics[width=0.8\columnwidth]{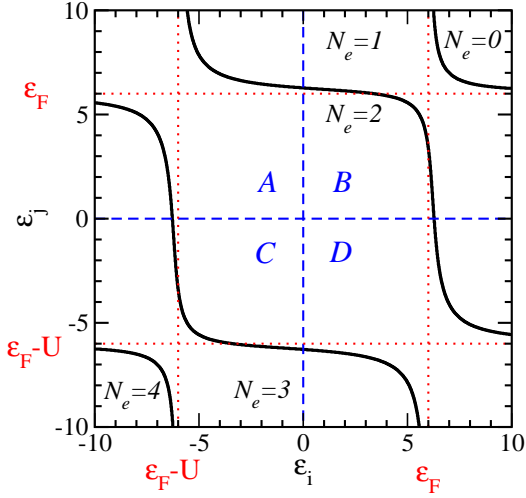}
\end{center}
\caption{Phase diagram for an isolated pair $(i,j)$ of sites with site
  energies $\epsilon_i$ and $\epsilon_j$.  The figure shows the number
  $N_e$ of electrons in the ground state, and is divided into four
  quadrants by the dashed blue lines. The quadrants are labelled
  A,\ldots,D.  For the 2-electron ground state, the most important
  regions are $\epsilon_i\sim\epsilon_j+U\sim \ef$ (quadrant D) and
  $\epsilon_i+U\sim \epsilon_j\sim \ef$ (quadrant A), which correspond
  to the LHO on one site and the UHO on the other being nearly
  degenerate with $\ef$.  For the 1-electron and 3-electron ground
  states, the most important regions of the phase diagram are
  $\epsilon_i \sim \epsilon_j \sim \ef$ (quadrant B) and $\epsilon_i
  \sim \epsilon_j \sim \ef-U$ (quadrant C) respectively.}
\label{fig:phasediag}
\end{figure}
 
Here, we calculate the derivative $\partial
\Lambda^\dprime_\mathrm{LHO}(E)$ for an ensemble of pairs $(i,j)$ of
isolated sites with random energies.
The Green's function for site $j$ with site $i$ removed is the atomic Green's function 
\begin{eqnarray}
G^\slashi_{jj}(\omega) &=& \frac{h_{j\sigmab}}{\omega-\epsilon_j} +
\frac{n_{j\sigmab}}{\omega-\epsilon_j-U}, \label{a:aG1}\\
&=& \frac{\omega-\epsilon_j - Uh_{j\sigmab}}{(\omega-\epsilon_j)(\omega-\epsilon_j-U)},
\label{a:aG2}
\end{eqnarray}
where $\sigmab=-\sigma$, $h_{j\sigmab}=1-n_{j\sigmab}$, and where we
suppress the spin index on $G$ in the nonmangnetic state.
\begin{widetext}
Using Eq.~(\ref{a:aG2}), 
\begin{eqnarray}
\Lambda^\dprime_{\epsilon_i}(\omega)  &=& \frac{t^2U^4}{\Delta}\int_{-\Delta/2}^{\Delta/2} d\epsilon_j 
\frac{p_{ij}^2}{(\omega-\epsilon_i-Uh_{i\sigmab})^2(\omega-\epsilon_j-Uh_{j\sigmab})(\omega-\epsilon_j)(\omega-\epsilon_j-U)},
\label{eq:L00}
\end{eqnarray}
and 
\begin{eqnarray}
\left. \frac{\partial \Lambda_{\epsilon_i}^\dprime(\omega)}{\partial \epsilon_i}\right |_{\omega=\epsilon_i} &=& 
\frac{t^2U^2}{\Delta} \int_{-\Delta/2}^{\Delta/2} d\epsilon_j  \left [
\partial_{\epsilon_i}p_{ij}^2
-\frac{2p_{ij}^2(1+U\partial_{\epsilon_i}h_{i\sigmab})}{Uh_{i\sigmab}}
-\frac{p_{ij}^2U\partial_{\epsilon_i}h_{j\sigmab}}{\epsilon_j+Uh_{j\sigmab}-\epsilon_i}
\right ] \nonumber \\
&&\times\frac{-1}{h_{i\sigmab}^2(\epsilon_j+Uh_{j\sigmab}-\epsilon_i)(\epsilon_j-\epsilon_i)(\epsilon_j+U-\epsilon_i)}
\label{a:dL00}
\end{eqnarray}
As discussed in the main text, the term in the square brackets proportional
to $p_{ij}^2/Uh_{i\sigmab}$ is a factor $t/U$ smaller than the other
terms, and is discarded.
\end{widetext}

Each pair of sites in the ensemble may have anywhere from 0 to 4 electrons,
depending on $\epsilon_i$ and $\epsilon_j$, and in order to evaluate
the integral in Eq.~(\ref{a:dL00}), we need to keep track of the
different states.  Figure \ref{fig:phasediag} shows that there are
four different possible ground states when $\epsilon_i$ is fixed near
$\ef$, having a total of $0$,$1$, $2$, or $3$ electrons shared between
$i$ and $j$.  

The 0-electron ground state does not contribute to $\partial
\Lambda_\mathrm{LHO}(E)$ because $p_{ij}=0$.  The 3-electron case also
does not make a substantial contribution, in this case because the
derivatives $\partial_{\epsilon_i}h_{i\sigmab}$,
$\partial_{\epsilon_i}h_{j\sigmab}$, and $\partial_{\epsilon_i}p_{ij}$
are of order $t^2/U$.  This follows because, in region D of the phase
diagram (Fig.~\ref{fig:phasediag}), the ground state wavefunction has
the form
\[
|3e\rangle \approx |\sigma 2\rangle - \frac{t}{\epsilon_i-\epsilon_j}|2\sigma\rangle,
\]
 where $|\sigma 2\rangle$
indicates that site $i$ has a single spin-$\sigma$ electron and that
site $j$ is doubly occupied.  Because $\epsilon_i-\epsilon_j \sim U$,
the derivatives in Eq.~(\ref{a:dL00}) are of order $t/U^2$. 

The remaining contributions to Eq.~(\ref{a:dL00}) are from the
1-electron and 2-electron ground states.  It turns out that $p_{ij}^2$
is an order of magnitude smaller in the 1-electron case, where
$\langle \hat S_{i+}\hat S_{j-}\rangle=0$, than in the 2-electron
case, and so we focus our attention on the latter.

For $\epsilon_i \sim \ef$, the most important contributions to the ZBA
for the 2-electron case come from quadrant D of the phase diagram in
Fig.~\ref{fig:phasediag}.  In this quadrant, there are two important
configurations: the singlet $|s\rangle =
(|\uparrow\downarrow\rangle-|\downarrow\uparrow\rangle)/\sqrt{2}$, and
the double-occupancy state $|02\rangle$ which has both electrons on
site $j$.  These states have nearly the same energy, so we use
degenerate perturbation theory.  We project the AHM onto $|s\rangle$
and $|02\rangle$ to get the Hamiltonian matrix,
\begin{equation}
H_{2e} = \left [ \begin{array}{cc}
 \epsilon_i+\epsilon_j & -\sqrt{2}t \\
 -\sqrt{2}t & 2\epsilon_j + U 
 \end{array} \right ]
\end{equation}
from which follows the ground state wavefunction, $|2e\rangle = \alpha
|s\rangle + \sqrt{1-\alpha^2} |02\rangle$, with
\begin{equation}
\alpha^2 = \frac{1}{2} \left [ 1 + \frac{y}{\sqrt{y^2+2t^2}} \right ],
\end{equation}
where $y=(\epsilon_j+U-\epsilon_i)/2$.  We can write the expectation
values in Eq.~(\ref{a:dL00}) in terms of $\alpha^2$:
\begin{eqnarray}
h_{i\sigmab} &=& 1-\frac{\alpha^2}{2}, \quad
h_{j\sigmab} = \frac{\alpha^2}{2}, \\
p_{ij} &=& -\alpha^2\left (1- \frac{\alpha^2}{4} \right ) \approx -\alpha^2
\end{eqnarray}
It simplifies our calculations significantly that the derivatives in Eq.~(\ref{a:dL00}) all reduce
to derivatives of $\alpha^2$.
We use $\partial_{\epsilon_i} = -\frac 12 \partial_y$ and substitute $d\epsilon_j \rightarrow 2dy$ to obtain
\begin{equation}
\left. \frac{\partial \Lambda_{\epsilon_i}^\dprime(\omega)}{\partial \epsilon_i}\right |_{\omega=\epsilon_i} =
\frac{t^2}{\Delta} \int_{y_0}^{y_1} \frac{dy}{y} \alpha^2 (\partial_y \alpha^2) \frac{1+\frac{\alpha^2}{4}}{1-\frac{\alpha^2}{2}}.
\label{a:dL01}
\end{equation}
We need the principal part of the integral for $\partial
\Lambda^\dprime_\mathrm{LHO}(E)$.  In deriving this expression, we
have neglected terms of order $t/U$.

The integration limits are given by the range of $y$ over which the
ground state has two electrons.  It can be shown that the 2-electron
state is stable in region D of Fig.~\ref{fig:phasediag} for\cite{Hongyi2010b}
\begin{equation}
\tilde \epsilon_i (\tilde \epsilon_j+U) <  2t^2,
\label{eq:constraint}
\end{equation}
where $\tilde \epsilon_{i,j} = \epsilon_{i,j} -\ef$.  The boundaries
of the 2-electron phase in this region are shown as thick black lines
in the figure. Setting $\epsilon_i = E$, we obtain the integration
limits
\begin{eqnarray}
y_0 &=& \frac{t^2}{\tilde E} - \frac{\tilde E}{2}, \qquad y_1 = \infty, \qquad (\tilde E<0) \\
y_0 &=& -\infty, \qquad y_1 =  \frac{t^2}{\tilde E} - \frac{\tilde E}{2}, \qquad (\tilde E>0),
\end{eqnarray}
 where $\tilde E = E-\ef$.  The integration limits at $\pm \infty$
 come from the boundaries of region D, which are taken to be far from
 $\ef$ and $\ef-U$.  This assumption does not change our results
 significantly because the integrand in Eq.~(\ref{a:dL01}) is peaked
 near $y=0$, because of the factor
\[
\partial_y \alpha^2 = \frac{t^2}{(y^2+2t^2)^{3/2}}.
\]
  We, for the same reason, can expand 
\begin{eqnarray*}
\left ( 1-\frac{\alpha^2}{2} \right )^{-4} &=& \left (\frac 43\right )^4 \left (1-\frac{y}{3\sqrt{y^2+2t^2}} \right )^{-4}\\
&\approx &\left (\frac 43\right )^4 \Bigg [ 1 + \frac{4y}{3\sqrt{y^2+2t^2}} + \frac{10y^2}{9(y^2+2t^2)} \nonumber \\
&&+ \ldots\Bigg] 
\end{eqnarray*}
and
\begin{eqnarray*}
 1+\frac{\alpha^2}{4} &=& \frac 98 +\frac{y}{8\sqrt{y^2+2t^2}} \approx \frac 98 
\end{eqnarray*}
to obtain
\begin{eqnarray}
\partial\Lambda_\mathrm{LHO}^\dprime(E) &=& \left. \mbox{Re }\frac{\partial \Lambda_{\epsilon_i}^\dprime(\omega)}{\partial \epsilon_i}\right |_{\omega=\epsilon_i=E} \nonumber \\
&=& \frac{4\sqrt{2}t}{9\Delta}\left [ F_1\left( \frac{y}{\sqrt{2}t} \right ) +
\frac{7}{6}F_2\left( \frac{y}{\sqrt{2}t} \right ) \right .\nonumber \\
&&\left.  - \frac{22}{27} F_3\left( \frac{y}{\sqrt{2}t} \right ) 
+\frac{25}{54} F_4\left( \frac{y}{\sqrt{2}t} \right ) 
\ldots \right ]_{y_0}^{y_1}, \nonumber \\
\end{eqnarray}
where $F_n(x) = n\int dx \, {x^{n-2}}/{(1+x^2)^{1+n/2}}$.  The
functions $F_n(x)$ are odd (even) when $n$ is even (odd).  Because the
limits $y_0$ and $y_1$ are odd in $\tilde E$, $F_2(x)$ and $F_4(x)$
actually make a contribution to $\partial \Lambda_\mathrm{LHO}(E)$
that is even in $\tilde E$.  Using the symmetry of Eq.~(\ref{eq:sym}),
it follows that
\begin{eqnarray}
\partial\Lambda_\mathrm{LHO}^\dprime(E) + \partial\Lambda_\mathrm{UHO}^\dprime(E)
 &=& \frac{28\sqrt{2}t}{27\Delta}\left [
F_2\left ( \frac{y}{\sqrt{2}t} \right ) \right.
\nonumber \\
&& \left.+ \frac{25}{63} F_4\left ( \frac{y}{\sqrt{2}t} \right ) \right]_{y_0}^{y_1} \nonumber \\
\label{a:dosdprime}
\end{eqnarray}

  Explicitly,
\begin{eqnarray}
F_2(x) &=&  \frac{x}{x^2+1}+\tan^{-1}(x). \label{a:F2} \\
F_4(x)  &=& \frac 12 F_2(x) - \frac{x}{(1+x^2)^2} 
\end{eqnarray}


\section{Symmetries of $\partial \Lambda_\mathrm{LHO}(E)$ and $\partial \Lambda_\mathrm{UHO}(E)$}
\label{sec:B}
In this appendix, we prove the relation $\partial
\Lambda_\mathrm{UHO}( E) = \partial \Lambda_\mathrm{LHO}(-E)$ (for
convenience, we take $\ef=0$ in this section).  This result is based on the
symmetries of the single-site Green's function, Eq.~(\ref{a:aG1}), and
the self-energy Eq.~(\ref{eq:approxSE}).

The proof proceeds as follows: $\partial \Lambda_\mathrm{LHO}(E)$ has
contributions from 2-electron states in region D of
Fig.~\ref{fig:phasediag} and 1-electron states in region B; $\partial
\Lambda_\mathrm{UHO}(E)$ has contributions from 2-electron states in
region A of Fig.~\ref{fig:phasediag} and 3-electron states in region
C.  We show that there is a correspondence between regions A and
D, and between regions B and C, with the result that
$\partial_\epsilon \Lambda_\epsilon(E)|_{\epsilon=E}$ in B (or D) is
equal to $\partial_\epsilon \Lambda_\epsilon(-E)|_{\epsilon=E-U}$ in C
(or A).

Suppressing subscripts, we write $\partial_\epsilon \Lambda_\epsilon(E)$, with
$\Lambda_\epsilon(E)$ given by Eq.~(\ref{eq:lambda2}), as
\begin{equation}
\partial \Lambda = 2 \langle (-t+\Sigma)(\partial \Sigma) G \rangle + \langle (-t+\Sigma)^2 \partial G\rangle.
\label{a:plambda}
\end{equation}
Now consider a pair of sites with $\epsilon_i$ and $\epsilon_j$ belonging to region D, and  a corresponding pair of sites with $\epsilon_i^\prime = \epsilon_i - U$, $\epsilon_j^\prime = \epsilon_j + U$ belonging to region A.  For region D, the wavefunction is 
$|2e\rangle = \alpha_{y}|s\rangle + \sqrt{1-\alpha_y^2}|02\rangle$ with $y=(\epsilon_j+U-\epsilon_i)/2$, while for
region A, $|2e\rangle^\prime = \alpha_{y}|s\rangle + \sqrt{1-\alpha_{y}^2}|20\rangle$ with $y=(\epsilon_i+U-\epsilon_j)/2$, where $\alpha_{y}$ is the same in both cases.  Because of this symmetry, $h_{j\sigmab}^\prime =  n_{j\sigmab}$ and $n_{j\sigmab}^\prime =  h_{j\sigmab}$.  It follows immediately that
the local Green's function Eq.~(\ref{a:aG1}) satisfies
\begin{equation}
G^\prime_{\epsilon_i+U=E} = - G_{\epsilon_i=E}, \quad
\partial G^\prime_{\epsilon_i+U=E} = \partial G_{\epsilon_i=E}
\label{a:Gsym}
\end{equation}
where $G^\prime$ is the Green's function for primed site energies, and $G$ is for unprimed site energies.   It also follows that $\Sigma_{ij}(E)\rightarrow \Sigma(y)$
for regions A and D, with $\Sigma(y)$ the same even function of $y$ in both cases, but with $y$ specific to each region, as above.  Thus
\begin{equation}
\Sigma^\prime_{\epsilon_i=E-U} = \Sigma_{\epsilon_i=E}, \quad
\partial \Sigma^\prime_{\epsilon_i=E-U} = -\partial \Sigma_{\epsilon_i=E}.
\label{a:Ssym}
\end{equation}
Equations (\ref{a:Gsym}) and (\ref{a:Ssym}) suggest that $\partial \Lambda$ is even under $(\epsilon_i,\epsilon_j) \rightarrow (\epsilon_i^\prime,\epsilon_j^\prime)$; however, an additional negative sign arises from averaging over $j$.  For $\epsilon_i$,
\[
\int d\epsilon_j \ldots \rightarrow \int_{y_0}^{y_1} 2dy \ldots,
\]
while for $\epsilon_i^\prime$
\[
\int d\epsilon_j^\prime \ldots \rightarrow \int_{-y_1}^{-y_0} 2dy \ldots 
\]
Because of the inverted integration limits, we obtain (considering only
contributions from regions A and D),
\begin{equation}
\partial_\epsilon \Lambda_\epsilon(E)|_{\epsilon=E} = \partial_\epsilon \Lambda_\epsilon(-E)|_{\epsilon=E-U}.
\end{equation}
An identical result is found if we consider primed and unprimed site energies belonging to regions B and C respectively, which proves
\begin{equation}
\partial \Lambda_\mathrm{LHO}(E) = \partial \Lambda_\mathrm{UHO}(-E).
\label{a:Lsym}
\end{equation}



\end{document}